\documentclass[reprint, amsmath, amssymb, aps, pra, twocolumn, superscriptaddress]{revtex4-2}

\usepackage{graphicx}
\usepackage{dcolumn}
\usepackage{bm}
\usepackage{hyperref}
\hypersetup{colorlinks=true, citecolor=blue, urlcolor=blue, linkcolor=blue}

\begin{document}

\title{Preparation of quasi-two-dimensional Bose mixture of ultracold $^{23}$Na and $^{87}$Rb atoms}

\author{Ji-Kai Liao}
\affiliation{Hefei National Research Center for Physical Sciences at the Microscale and Department of Modern Physics, University of Science and Technology of China, Hefei 230026, China}
\affiliation{Shanghai Research Center for Quantum Science and CAS Center for Excellence in Quantum Information and Quantum Physics, University of Science and Technology of China, Shanghai 201315, China}
\author{Hao-Ran Zhang}
\affiliation{Hefei National Research Center for Physical Sciences at the Microscale and Department of Modern Physics, University of Science and Technology of China, Hefei 230026, China}
\affiliation{Shanghai Research Center for Quantum Science and CAS Center for Excellence in Quantum Information and Quantum Physics, University of Science and Technology of China, Shanghai 201315, China}
\author{Xiao-Rong Yu}
\affiliation{Hefei National Research Center for Physical Sciences at the Microscale and Department of Modern Physics, University of Science and Technology of China, Hefei 230026, China}
\affiliation{Shanghai Research Center for Quantum Science and CAS Center for Excellence in Quantum Information and Quantum Physics, University of Science and Technology of China, Shanghai 201315, China}
\author{Ya-Qun Qi}
\affiliation{Hefei National Research Center for Physical Sciences at the Microscale and Department of Modern Physics, University of Science and Technology of China, Hefei 230026, China}
\affiliation{Shanghai Research Center for Quantum Science and CAS Center for Excellence in Quantum Information and Quantum Physics, University of Science and Technology of China, Shanghai 201315, China}
\author{Yi-Cheng Guo}
\affiliation{Hefei National Research Center for Physical Sciences at the Microscale and Department of Modern Physics, University of Science and Technology of China, Hefei 230026, China}
\affiliation{Shanghai Research Center for Quantum Science and CAS Center for Excellence in Quantum Information and Quantum Physics, University of Science and Technology of China, Shanghai 201315, China}
\author{Bo Zhao}
\affiliation{Hefei National Research Center for Physical Sciences at the Microscale and Department of Modern Physics, University of Science and Technology of China, Hefei 230026, China}
\affiliation{Shanghai Research Center for Quantum Science and CAS Center for Excellence in Quantum Information and Quantum Physics, University of Science and Technology of China, Shanghai 201315, China}
\affiliation{Hefei National Laboratory, University of Science and Technology of China, Hefei 230088, China}
\author{Jun Rui}
\affiliation{Hefei National Research Center for Physical Sciences at the Microscale and Department of Modern Physics, University of Science and Technology of China, Hefei 230026, China}
\affiliation{Shanghai Research Center for Quantum Science and CAS Center for Excellence in Quantum Information and Quantum Physics, University of Science and Technology of China, Shanghai 201315, China}
\affiliation{Hefei National Laboratory, University of Science and Technology of China, Hefei 230088, China}
\author{Jian-Wei Pan}
\affiliation{Hefei National Research Center for Physical Sciences at the Microscale and Department of Modern Physics, University of Science and Technology of China, Hefei 230026, China}
\affiliation{Shanghai Research Center for Quantum Science and CAS Center for Excellence in Quantum Information and Quantum Physics, University of Science and Technology of China, Shanghai 201315, China}
\affiliation{Hefei National Laboratory, University of Science and Technology of China, Hefei 230088, China}

\date{\today}

\begin{abstract}
Quantum gases confined in reduced dimensions have enabled the observation of many exotic quantum phenomena. While existing experiments primarily focus on homonuclear systems, we report here on the efficient preparation of a quasi-two-dimensional (2D) heteronuclear quantum degenerate mixture of ultracold $^{23}$Na and $^{87}$Rb. We describe the apparatus design and detail the procedures for preparing the 2D quantum mixture. The new apparatus has several unique features, including compact and modular 2D-MOT sources, a science chamber that accommodates various lattice geometries, a precision in-vacuum electrode assembly, and high-resolution imaging for both atomic species. After loading the dual-species condensate into a single layer of a vertical optical lattice, we prepare a 2D gas mixture and observe quantum immiscibility in the \textit{in situ} equilibrium density profiles. The observed density profiles reasonably agree with mean-field theories. The apparatus provides a versatile platform for investigating several interesting problems, including quantum impurities, quantum droplets, or polar molecules in low dimensions.

\end{abstract}

\maketitle

\section{Introduction}

Ultracold atomic mixtures offer independent control and measurement over constituent species and tunable interspecies interactions, emerging as a powerful platform for exploring complex quantum few- and many-body physics \cite{bloch_RMP_2008, chin_fesbach_2010, baroni_mixtures_2024}. Research in this field spans a broad range of fundamental questions, including tunable quantum miscibility~\cite{Papp_Rb-mixture_2008, Naidon_bose-mixture_2021, Jia_shell_2022}, the formation of quantum droplets~\cite{Petrov_droplet_2015, Cabrera_droplet_2018, Semeghini_droplet_2018, derrico_krb_2019, Skov_droplet_2021}, and the properties of Fermi or Bose polarons \cite{schirotzek_polaron_2009, jorgensen_polaron_2016, hu_polaron_2016, Zoe_polaron_2020}, etc. These mixtures further enable the association and production of ultracold quantum degenerate polar molecules in recent years~\cite{valtolina_molecule_2020, Schindewolf_molecule_2022, bigagli_NaCs-BEC_2024, Shi_NaRb-BEC_2025}. A wide array of dual-species mixture experiments - utilizing combinations such as $^{85}$Rb--$^{87}$Rb \cite{Papp_Rb-mixture_2008}, $^{6}$Li--$^{40}$K \cite{Costa_LiK_2010}, $^{6}$Li--$^{133}$Cs \cite{Pires_LiCs_2014}, $^{23}$Na--$^{40,39}$K \cite{Park_NaK_2012, Zhu_NaK_2017, Schulze_39KNa_2018},  $^{23}$Na--$^{87}$Rb \cite{wang_bec_2016, Jia_shell_2022,rosenberg_molecule_2022}, $^{23}$Na--$^{133}$Cs \cite{Warner_NaCs_2021}, $^{39,40,41}$K--$^{87}$Rb ~\cite{Wacker_39K87Rb_2015, Modugno_K-Rb_2002, derrico_krb_2019} - have demonstrated the versatility of heteronuclear systems. While these studies mainly focus on tunable interactions in three dimensions (3D), tight geometrical confinement of quantum gases into lower dimensions introduces an additional indispensable degree of freedom that enables the observation of unique critical behaviors, most notably the Berezinskii-Kosterlitz–Thouless (BKT) phase transition \cite{hadzibabic_bkt_2006, Hung_2d-invaraince_2011}. Previous experiments have successfully prepared 2D mixtures of $^{40}$K--$^{87}$Rb \cite{valtolina_molecule_2020} and $^{23}$Na--$^{87}$Rb \cite{rosenberg_molecule_2022}. Nevertheless, these works focused primarily on preparing ultracold polar molecules in 2D and did not explore phenomena in quasi-2D atom mixtures.

In this work, we extend these capabilities by preparing a heteronuclear Bosonic mixture in the BKT transition regime for the first time. For ultracold quantum mixtures confined to low dimensions ~\cite{Bakkali-2d-mix-2021}, a variety of novel quantum phases are predicted to emerge from the interplay between enhanced quantum fluctuations and tunable short- or long-range interactions. For instance, balanced mixtures may lead to the formation of a stable quantum droplet in 2D as long as the interspecies interaction is attractive \cite{Petrov_droplet_2016}. Interestingly, a mixture in the gas phase in 3D may transition to a liquid state upon confinement to a quasi-2D geometry \cite{Spada_droplet_2024}. Furthermore, if an atomic mixture is assembled into polar molecules and confined to a 2D geometry, collisional shielding may be engineered via a DC electric or microwave field \cite{Micheli_2dcollsion_2007, Gorshkov_shield_2008}. Given sufficient collisional stability, the long-range interactions may lead to the formation of quantum crystals in 2D pancake traps \cite{buechler_molphase_2007} or novel quantum phases in 2D optical lattices \cite{capogrosso_mol_2dlattice_2010}. To enable microscopic study of these phenomena, an advanced apparatus producing ultracold mixtures and offering versatile optical, magnetic, and electric-field controls - particularly capable of single-site detection and addressing in various lattice geometries \cite{bakr_microscope_2009, sherson_microscope_2010, rosenberg_molecule_2022} - is highly desired and challenging. 

\begin{figure*}[t]
	\includegraphics[width=145mm]{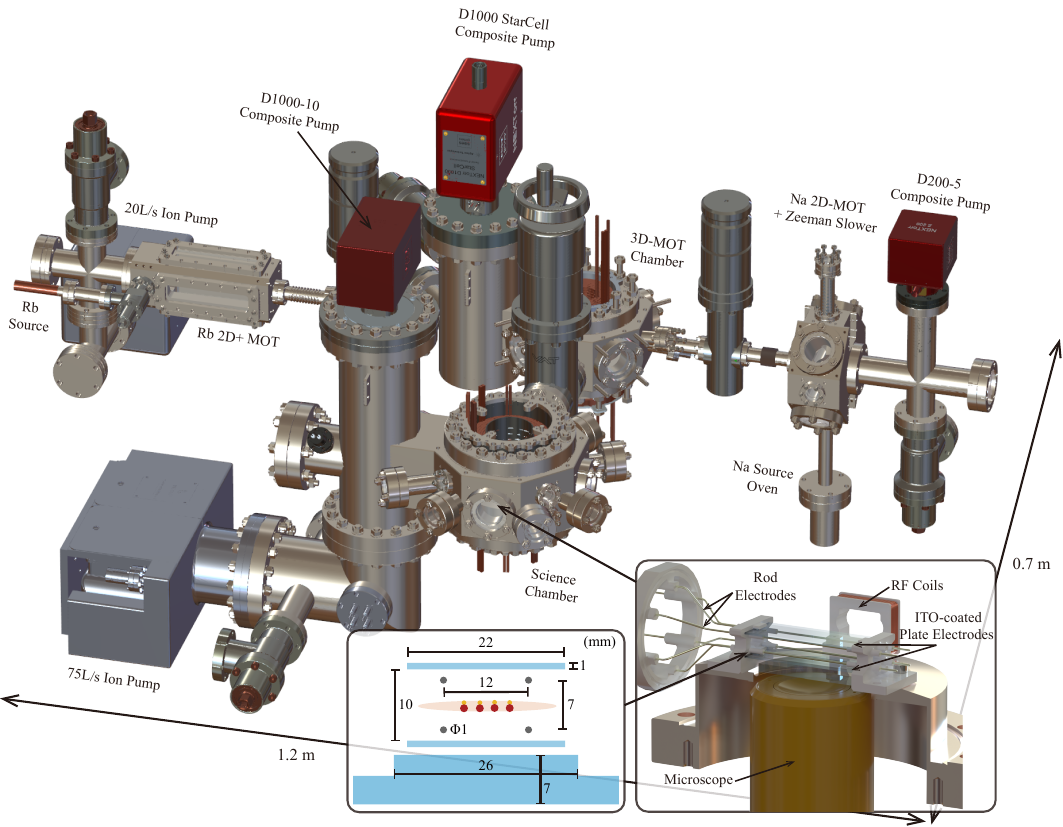}
	\caption{\label{fig:vac} Schematic of the vacuum system. The apparatus comprises four primary sections: a $^{23}$Na 2D-MOT chamber, a $^{87}$Rb 2D-MOT chamber, a 3D-MOT chamber, and a science chamber. The two 2D-MOT chambers are connected to the 3D-MOT chamber via differential pumping tubes, bellows, and CF16 all-metal, non-magnetic gate valves. The chamber of the Rb 2D-MOT is indium-sealed. The 3D-MOT chamber is connected to the science chamber through a CF35 gate valve and a CF35 bellow. (Inset) The detailed view highlights the in-vacuum Kapton-insulated RF coils wound on a Macor support, the precision electrode assembly mounted on the end plate of the bottom CF125 re-entrant window, and the high-resolution imaging objective with a numerical aperture (NA) of 0.75. The sketch beside displays the geometry and dimensions of the electrode system and the bottom window glass. }
\end{figure*}

This paper describes the construction of an experimental apparatus for preparing quasi-2D $^{23}$Na--$^{87}$Rb mixtures. This apparatus features a modular vacuum and optical design for laser cooling that creates large atom fluxes of both atomic species, and a science chamber with several key capabilities: a viewport layout that accommodates various lattice geometries, the stable alignment of a precision in-vacuum electrode assembly, high-current, water-cooled magnetic coils, species-dependent loading of the mixture in bichromatic vertical lattices, and high-resolution imaging of both atomic species. After a series of procedures, the mixture is efficiently loaded into a single layer of a vertical optical lattice with a large lattice constant. Finally, we observe the immiscible \textit{in situ} density distribution of quasi-2D Bose mixture via high-resolution absorption imaging. 

This paper is organized as follows. Section II provides an overview of the apparatus and highlights the main features. Section III describes the implementation of optical transport into the Science chamber and optical evaporation. Section IV details the preparation of the quasi-2D quantum gas, and Section V describes the observation of immiscibility in the quasi-2D Bose mixture. The last section presents conclusions and outlines possible future directions. 

\section{Overview of the Apparatus}

Our setup utilizes a compact Zeeman slower combined with a 2D-MOT for $^{23}$Na atoms \cite{lamporesi_na2dmot_2013} and a newly designed modular 2D$^{+}$-MOT for $^{87}$Rb atoms \cite{dieckmann_2dmot_1998}, both employing permanent magnets to provide high-flux slowed atomic beams, as illustrated in Fig. \ref{fig:vac}. Centered between these sources is an octagon 3D-MOT chamber, featuring CF100 re-entrant windows for the installation of high-gradient magnetic coils. To maximize pumping speed and maintain ultra-high vacuum (UHV), a CF100 pumping tube is welded directly to the octagon, with a getter pump mounted. This 3D-MOT chamber serves as the primary stage for capturing laser-cooled atoms, performing molasses cooling, and implementing forced evaporation within a magnetic trap. As the discussion of these procedures has already been covered in many studies, the details of the designs and parameters are provided in the Appendix. Most vacuum components were manufactured from non-magnetic TC4 titanium alloy, except for a few commercial parts (e.g., pumps and valves) made of stainless steel, thereby significantly suppressing eddy-current effects during high-magnetic-field switching off. With the large fluxes of both atom species, we are able to optimally load up to $2\times 10^9$ $^{87}$Rb atoms and $1\times10^{9}$ $^{23}$Na atoms per second in the 3D-MOT. Enabled by the large atom number of both species, we used a single frequency-swept RF signal to simultaneously evaporate atoms of both species in the magnetic trap (see more details in Appendix \ref{app-evaporation}).

\begin{figure}[h]
	\includegraphics[width=75mm]{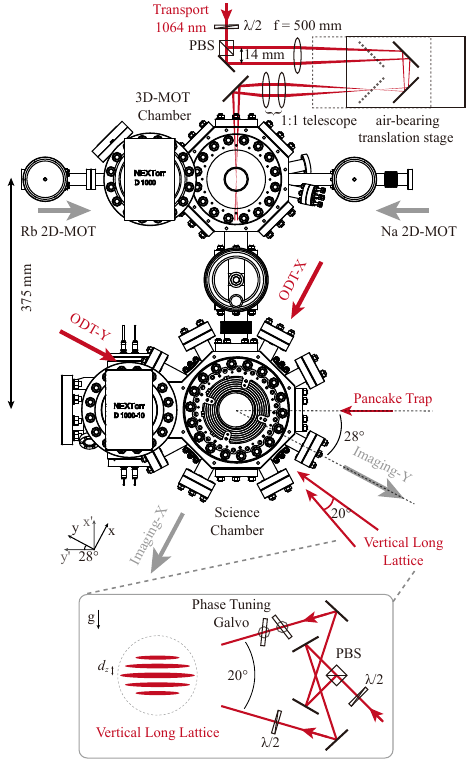}
	\caption{\label{fig:sciopt} Top view of the science chamber and the overall optical layout. A small-angle crossed ODT is formed by focusing two parallel 1064 nm laser beams with orthogonal polarizations. Its focal point can be translated by an air-bearing stage, as described in the main text, to transport atoms into the crossed ODT trap within the science chamber. A long vertical lattice is created by two interfering 1064 nm beams incident through upper and lower CF16 viewports at a crossing angle of 20$^{\circ}$. The two beams are matched in optical path length to ensure passive phase stability of the vertical long-lattice. In addition, the lattice phase can be actively controlled using a pair of scanning Galvos. } 
\end{figure}

To overcome the optical and spatial constraints of the 3D-MOT chamber, we designed a dedicated science chamber with enhanced optical access. The science chamber is equipped with eight CF25 and eight tilted CF16 vacuum windows, enabling a variety of horizontal lattice geometries and a vertical long lattice (VLL) formed by beams intersecting at $\pm 10^{\circ}$ relative to the horizontal plane, as shown in Fig. \ref{fig:sciopt}. Additionally, two CF125 re-entrant windows enable high-resolution imaging along the vertical direction and the formation of a vertical short-lattice (VSL) via retroreflection from the bottom window. Together, these vertical lattice configurations provide variable confinement for studying 2D atomic or molecular physics in the future. 

To associate Feshbach molecules and tune the dipolar interactions of ground-state $^{23}$Na$^{87}$Rb molecules in the future, the science chamber houses a set of high-current magnetic coils and a precision in-vacuum electrode assembly ~\cite{covey_phd_2017,valtolina_molecule_2020}, as shown in the inset of Fig. \ref{fig:vac}. The magnetic coils can produce a homogeneous field for the Feshbach resonance at around 347.6 G \cite{wang_bec_2016} with a driving current of about 100 A . The in-vacuum electrode assembly consists of a pair of ITO-coated thin glass plates (62 mm × 22 mm × 1 mm) placed parallel, separated by 10 mm, to provide a uniform DC electric field. In addition, four tungsten rods (1 mm diameter) are arranged in a rectangular configuration spanning 12 mm × 7 mm to enable versatile controls of the electric field ~\cite{covey_phd_2017}. The entire assembly is mounted on the endplate of the bottom CF125 re-entrant window. An RF coil wound on a Macor support with kapton-insulated copper wire is also mounted on the same endplate for the Zeeman sublevel transfer of atoms at low magnetic fields. Because the lower ITO-coated glass plate is positioned within the high-resolution imaging path, its installation and alignment are critical. The direct-mounting strategy ensures the stable alignment of the electrodes relative to the vacuum window throughout the vacuum pumping and high-temperature baking. The specific procedures used to ensure diffraction-limited optical performance will be detailed in a separate publication. With all the electrodes and wires installed, the background pressures in both the 3D-MOT and science chambers are maintained below $1\times 10^{-11}$ mbar during normal operations. 

To probe the \textit{in situ} density profiles of the 2D gases and the effects of interspecies interactions, our apparatus is designed to support high-resolution imaging of both atomic species. The optical objective features an NA of 0.75 and an effective focal length (EFL) of 12 mm, as shown in Fig.\ref{fig:vac}. The working distance accounts for 8 mm of fused-silica glass (consisting of a 1 mm electrode plate and a 7 mm vacuum window), 6 mm in vacuum, and 1 mm in air. While the primary tube lens has an EFL of $650$ mm, providing a magnification of $\approx54$ for single-site resolved detection, we inserted an achromatic lens ($f=100$ mm) after the tube lens to reduce the effective magnification to $\approx$12 for this measurement. Both the objective and the tube lens have been optimized for performance at the imaging wavelengths of both species.

\section{Optical Transport and Evaporation}

With the near-degenerate mixture prepared in the magnetic trap in the 3D-MOT chamber (Appendix \ref{app-evaporation}), the atoms were transferred to the science chamber via a transport optical dipole trap (ODT). We used a crossed optical trap with a small angle of 1.6$^{\circ}$ and a waist of 55 $\mu$m to enhance its axial confinement \cite{gross_transport_2016}, as shown in Fig. \ref{fig:sciopt}. A 1:1 telescope composed of two $f=400$ mm lenses relayed the focused crossed beams into the vacuum chamber. By moving the air-bearing stage, the intersection point of the optical trap could be shifted 375 mm from the center of the 3D-MOT chamber to the center of the science chamber. The total trap power was about 13.2 W, and each ODT beam provided a depth of 56 $\mu$K for Na and 184 $\mu$K for Rb. In this transport trap, the measured trap frequencies were $\omega_{\text{axial}}$ = 2$\pi$× 18.6 (18.1) Hz and $\omega_{\text{radial}}$ = 2$\pi$× 1169 (1121) Hz for Na (Rb).

During the transport, we observed a vector field for Rb corresponding to a magnetic field of 75 mG, pointing along the trap beam's propagation direction, using microwave spectroscopy. We attribute this effect to vector light shifts induced by the ODT beam's random elliptical polarization, arising from interference between the two perpendicular polarization light fields. As a result, we applied a small bias magnetic field during transport to maintain the quantization axis. The transfer trajectory follows an S-curve motion profile — a segmented constant-jerk motion comprising acceleration followed by deceleration \cite{gross_transport_2016}, which is adiabatically smooth to avoid heating. After optimization, the maximum jerk was set to 5000 mm/s$^3$, and the entire transfer process took about 1.1 s. With optimized RF evaporation and transport parameters, we routinely transfer up to 1.1×10$^7$ Na atoms and 1.0×10$^6$ Rb atoms to the science chamber. The overall transport efficiency exceeded 85\% for Na (whose trap depth is about 0.31 of that for Rb) and nearly 100\% for Rb, and the temperature of both atomic species remains 12.8 $\mu$K throughout the transfer. 

Following the optical transport, two orthogonally crossed ODTs (ODT-X/Y; see Fig. \ref{fig:sciopt}) were slowly turned on simultaneously, with the transport ODT power ramped down and then switched off. The crossed-ODT beams were designed to have elliptical profiles, with vertical and horizontal waists of 45 $\mu$m and 155 $\mu$m, respectively. To avoid interference, the two beams originate from separate lasers with a large frequency difference and orthogonal polarizations. The retro-reflections of the ODT-X/Y beams produce horizontal lattices for future experiments and fluorescence imaging. During the loading of the crossed ODT, however, the retro-reflection paths were blocked by a motorized blade. To efficiently load the mixture into a quasi-2D geometry, we introduced a pancake ODT to vertically compress the cloud. This trap, formed by a 1064 nm beam, has waists of 170 $\mu$m in the horizontal plane and 14 $\mu$m in the vertical direction. 

\begin{figure}[t]
	\includegraphics[width=80mm]{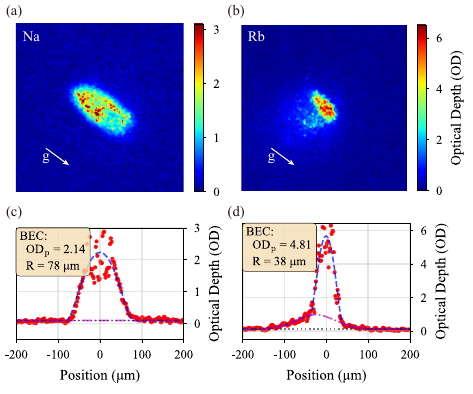}
	\caption{\label{fig:bec} Absorption images of dual-species 3D BECs after 15 ms TOF in the science chamber: (a) Na and (b) Rb. Imaging was performed along the horizontal X/Y directions. (c), (d) Horizontal cross-sections through the center of the 2D density distributions for Na and Rb, respectively. Blue dashed, purple dash-dotted, and gray dotted lines correspond to fits using the Thomas-Fermi distribution, the thermal cloud distribution, and the background, respectively. The density depletion in the Na cloud results from the interspecies immiscibility.}
\end{figure}

To load atoms from the transport ODT, both crossed-ODT beams were raised to 8 W, and the pancake trap to 440 mW simultaneously, leading to transfer efficiencies of about 55\% for Na and about 90\% for Rb. Evaporative cooling in this combined optical trap lasts over 6 s, with all optical powers being exponentially reduced using a 1.25 s time constant. At the end of the forced evaporation, the power of each crossed-ODT beam was lowered to 0.1 W and that of the pancake trap to 35 mW, yielding a dual-species BEC. The measured trap frequencies of Na (Rb) atoms were 2$\pi$×[28, 25, 283] ([26, 23, 263]) Hz. The BEC contained  2.1×10$^5$ Na atoms and 7.4×10$^4$ Rb atoms. The Na cloud was essentially pure BEC, while the Rb cloud had a condensed fraction of 49\% (see Fig.\ref{fig:bec}).

One can clearly observe phase separation between the two species in the density profiles of the 3D BECs after time-of-flight, a phenomenon that has already been studied in an earlier experiment \cite{wang_bec_2016}. The immiscibility of a two-component BEC is governed by the competition between intraspecies and interspecies interaction energies. In our experiment at nearly zero magnetic field, the relevant background 3D scattering lengths are $a_{11}=54.5\,a_{0}$ for $^{23}$Na \cite{knoop_scatter_2011}, $a_{22}=100.4\,a_{0}$ for $^{87}$Rb \cite{marte_scatter_2002}, and $a_{12}=76.3\,a_{0}$ for the Na-Rb mixture \cite{guo_scatter_2022}, with $a_{0}$ the Bohr radius. The interaction parameters satisfy $g_{12}^{2}>g_{11}g_{22}$, with $g_{ij}=2\pi\hbar^{2}a_{ij}/m_{ij}$ the 3D coupling constant and $m_{ij}$ the reduced mass. It thus drives the mixture into the immiscible regime in 3D, and results in partial density depletion in the Na BEC cloud that is expelled by the more tightly confined Rb component, as shown in Fig.\ref{fig:bec}.

\section{Preparing Quasi-2D Quantum Gas}

After producing dual-species condensates in the crossed-ODT, the atoms were loaded into a vertical long-lattice (VLL) to form a quasi-2D quantum-degenerate mixture \cite{dai_vll_2016}. The VLL is formed by equal optical-path interference of two 1064 nm laser beams of the same frequency crossing at $\theta = 20^{\circ}$ in the vertical plane. The two beams enter the chamber through the upper and lower CF16 viewports, as illustrated in Fig. \ref{fig:sciopt}. A common lens group shapes the beams and focuses them at the chamber center, with horizontal and vertical waists of 320 $\mu$m and 57 $\mu$m, respectively, resulting in nearly circular confinement in the 2D plane. The beams are horizontally polarized, and their interference produces an optical lattice with period $d_z = \frac{\lambda}{ 2 \sin(\theta/2)} \approx 3 \mu$m. A pair of scanning Galvos in a single beam path allows adjustment of the lattice phase \cite{Li_galvo_2021}, enabling precise loading of atoms into a single layer and alignment with the focal plane of the high-resolution imaging system. To characterize the passive phase stability of the long lattice, we magnified the interference pattern by a factor of 19 and recorded its intensity pattern with a small-pixel camera in an offline setup. We observed a peak-to-peak phase drift of about 0.23$\pi$ for a temperature change of $\sim$0.2 $^{\circ}\text{C}$/h. This stability meets the requirements for high-resolution imaging and the future transfer of the 2D gas into the VSL with a 532 nm lattice constant to achieve tighter vertical confinement. Complementing the 1064 nm VLL, we have integrated an additional species-dependent VLL using a 671 nm laser field to engineer the vertical relative wavefunctions of the two species. Although not utilized in the present study, these two lattice beams share a common beam path to ensure high relative stability. Furthermore, the relative phase between the two lattices can be precisely tuned by adjusting both Galvos, providing a versatile tool for future investigations of lattice-based quantum mixtures.  

In our experiment, after evaporative cooling in the combined optical trap, the vertical Thomas-Fermi radius of the trapped Na BEC was 1.6 $\mu$m. We then adiabatically increased the power of the pancake trap to 105 mW over 200 ms, further compressing the vertical radius of Na to 1.2 $\mu$m. For Rb, the smaller number of atoms results in a much tighter BEC, effectively precluding double-layer occupation. The total VLL laser power was increased to 406 mW within 20 ms, after which the crossed-ODT was adiabatically switched off, transferring all the atoms into a single layer of the VLL. At this stage, the measured trap frequencies for Na (Rb) in the single-layer VLL were 2$\pi\times[21.9,17.3,4390] ~([20.2,16.1,3980])$Hz. To verify single-layer loading, we exploited the condensate's coherence across different lattice layers and observed matter-wave interference after release from the lattice. If the condensate occupies multiple layers, the expanding clouds interfere, producing fringes after TOF whose period is related to the lattice spacing. The fringe period can be understood as the de Broglie wavelength associated with the relative motion of atoms $\lambda=\frac{ht}{md}$, where $t$ denotes the TOF duration, $m$ the atomic mass, and $d$ the separation between atomic clouds \cite{andrews_interference_1997}. From the measured fringe, the lattice spacing is estimated to be 3.0 $\mu$m. As shown in Fig. \ref{fig:2d}(a)\&(b), by adjusting the scanning Galvos, we confirmed that both Na and Rb atoms were loaded into a single lattice layer, as evidenced by the absence of interference fringes. 

\section{Immiscible Quasi-2D Bose Mixture}

\begin{figure}[t]
	\includegraphics[width=80mm]{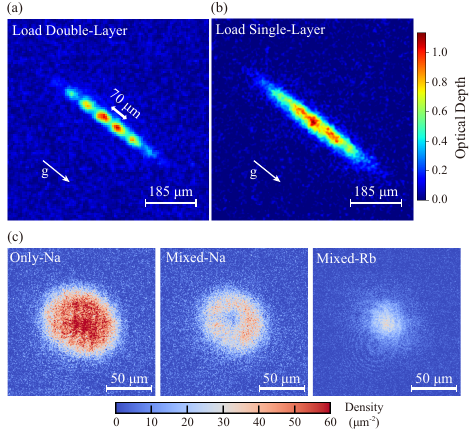}
	\caption{\label{fig:2d}Preparing the quasi-2D quantum mixture gas in the immiscible regime. (a) Matter-wave interference of a Na condensate loaded into two adjacent vertical long-lattice layers, with absorption imaging along the y-direction after 12 ms TOF. (b) Adjusting the galvanometer voltage to alter the lattice phase loads atoms into a single long-lattice layer, as confirmed by the disappearance of matter-wave interference. (c) \textit{In situ} radial density distribution of the trapped quasi-2D single-species and dual-species quantum gases. }
\end{figure}

In the experiment, we measured the \textit{in situ} 2D equilibrium density distribution of the 2D gases using short, high-intensity probe pulses through the high-resolution imaging setup \cite{yefsah_2d_2011}. In the single-species case, the equilibrium density distribution of thermal atoms in the low-density wing of a quasi-2D system is well described by the Hartree-Fock mean-field (HFMF) theory \cite{holzmann_2dtheory_2008,bisset_2dtheory_2009,roy_2d-bose-mixture_2021}, which can be calculated by the self-consistent theory. The atomic density thus takes the explicit form,
\begin{equation}
\begin{aligned}
n_i(r)\lambda^2_{\text{dB}i} &= \sum_\nu-\ln\left\{1- Z_\nu\right\},\\
Z_\nu &= \exp\left[\frac{\mu_i(r) -\nu\hbar\omega^z_i- 2g_{ii}^{2\text{D}}n_i(r)}{k_B T}\right].
\end{aligned}
\end{equation}

Here, $n_i(r)$ represents the local density of species $i$, while $\lambda_{\mathrm{dB}i}=h/\sqrt{2\pi m_i k_B T}$ denotes the thermal de Broglie wavelength, where $m_i$ is the mass and $T$ is the temperature. And $\omega^z_i$ is the trap frequency along the tightly confined z-axis, and $\nu$ is the motional quantum number with a cutoff value. The local chemical potential is defined as $\mu_i(r) = \mu_i - V_i(r)$, which incorporates the radial harmonic trapping potential $V_i(r)$ and the global chemical potential $\mu_i$ of the corresponding atomic species. The inter- and intra-species 2D coupling constant reads,
\begin{equation}
g_{ij}^{\rm 2D} = g_{ij} \sqrt{\frac{1}{\pi\hbar}\cdot \frac{m_i\omega^z_i \, m_j\omega^z_j}{m_i\omega^z_i + m_j\omega^z_j}}.
\end{equation} 

However, the HFMF theory is no longer applicable in the central high-density region dominated by the superfluid component. This region corresponds to local reduced chemical potential $\tilde{\mu}_i(r)= \mu_i(r)/k_B T > \tilde{\mu}_c$, with $\tilde{\mu}_c = (\tilde{g}/\pi)\ln(13.2/\tilde{g}) \approx0.082$ the critical chemical potential \cite{Prokofev_2dcritcal_2001} and $\tilde{g} = m_i g^{2D}_{ii} /\hbar^2\approx 0.046$ the dimensionless 2D interaction strength of the Na atoms. The density profile of this region is described by the mean-field (MF) theory \cite{Prokofev_2dtheory_2002,Hung_2d-invaraince_2011},
\begin{equation}
    n_i(r)\lambda^2_{\text{dB}i} = \frac{2\pi\tilde{\mu}_i(r)}{g_{ii}^{2\text{D}}} + \ln\left( \frac{2n_i\lambda^2_{\text{dB}i} g_{ii}^{2\text{D}}}{\pi} - 2\tilde{\mu}_i(r) \right),
\end{equation}
where the first term on the right-hand side describes the superfluid component and the second term describes the residual thermal cloud part. In the fluctuation region with $-1 <\frac{\tilde{\mu}-\tilde{\mu}_c}{\tilde{g}} < 0$, the 2d gas transitions from a thermal gas into a superfluid.

\begin{figure}[t]
	\includegraphics[width=80mm]{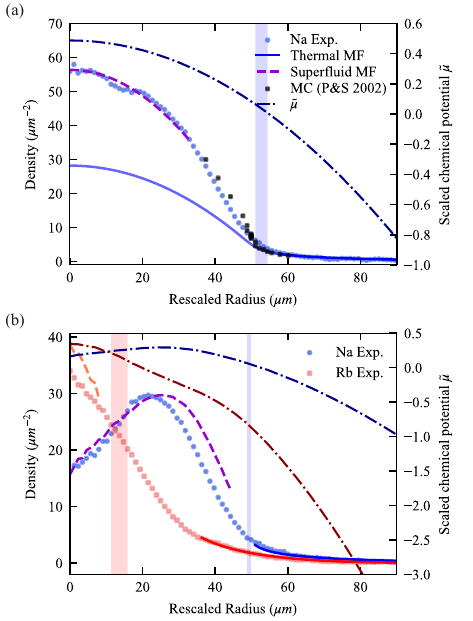}
	\caption{\label{fig:2drad} Azimuthally averaged \textit{in situ} 2D density profiles of (a) a single Na species and (b) a Na+Rb mixture. Blue (red) points represent the measured density of Na (Rb). Thermal wings are fitted with HFMF theory ($\tilde{\mu}<0$) to extract the temperature and chemical potential, and the theoretical density curves are shown as the blue (red) solid line for Na (Rb) atoms. Regions with $\tilde{\mu}>\tilde{\mu}_c$ are treated with superfluid MF theory, and the theoretical curve is shown as a purple (orange) dashed line for Na (Rb) atoms. (a) Both the HFMF and superfluid MF predictions agree well with the experiment, and we extend the HFMF curve into the superfluid region for comparison. The Monte Carlo result from Ref. \cite{Prokofev_2dtheory_2002} is also presented for comparison. (b) For the mixture, the radial density profiles clearly show 2D phase separation. And the superfluid MF theory agrees reasonably well with the measured density profiles.}
\end{figure}

The observed \textit{in situ} 2D equilibrium density profiles of a single Na species are presented in  Fig. \ref{fig:2d} (c) and Fig. \ref{fig:2drad} (a). With local density approximation, we perform an average over points with the same local chemical potential $\mu(r)$ to obtain the radial density distribution, taking into account the ellipticity in the radial trapping potential. We use the HFMF theory to fit the outer thermal wing with a local chemical potential $\tilde{\mu} <0$. To account for the ellipticity in the radial trap, we adopt the rescaled radius $R = \sqrt{x_e^2+\frac{\omega_{\max}^2}{\omega_{\min}^2}y_e^2}$, where $\omega_{\max}(\omega_{\min})$ is the max(min) radial trap frequency and $y_e(x_e)$ is the coordinate along the minor/major axis of the elliptical density distribution. The number of Na atoms is measured to be $2.5 \times 10^5$, and the fit yields a temperature of 111.5 nK and a chemical potential of $k_B \times 54.4$ nK. For the region with $\tilde{\mu}>\tilde{\mu}_c$, the measured density profile agrees well with the superfluid mean-field theory. In this regime, we record a peak phase-space density (PSD) of 69, which far exceeds the critical PSD for the BKT transition $\Lambda_c = \ln(380/\tilde{g})\approx9$ \cite{Prokofev_2dcritcal_2001}. For strictly 2D trapped Bose gas, the BKT temperature can be related to its ideal 2D BEC transition temperature \cite{Holzmann_2008},
\begin{equation}
T^{\rm 2d}_{\text{BKT}}/T^{\rm 2d}_\mathrm{BEC} =\left(1 + \frac {3\tilde {g}}{\pi^3} \Lambda_c^2\right)^{-\frac{1}{2}}\approx 0.86.
\end{equation}
Here the 2D BEC critical temperature is $T^{\rm 2d}_\mathrm {BEC}  = \sqrt {6N} \hbar\bar{\omega}_r / \pi k_\mathrm {B} \approx 364$ nK in our case, with $\bar{\omega}_r$ mean radial trap frequency. Taking into account the quasi-2D confinement, $\hbar\omega_z/T^{\rm 2d}_\mathrm{BEC}\approx 0.58$, one expects a quasi-2D BKT temperature to be slightly lower than the strictly 2D critical value $T^{\rm 2d}_{\text{BKT}}\approx 312$ nK for the 2d trapped Na atoms \cite{holzmann_2dtheory_2008}.  Nevertheless, this unambiguously indicates the emergence of a robust superfluid core at the trap center. 

In the quasi-2D Bose mixture case, the exact miscible-immiscible phase boundary is affected by many parameters, including mass ratios, chemical potentials, intraspecies interaction strengths, and temperature relative to the BKT transition point ~\cite{Karle_2d-bose-mixture_2019, Naidon_bose-mixture_2021, roy_2d-bose-mixture_2021}, etc. Nonetheless, using the criterion $g^{\text{2D}}_{12} = \sqrt{g^{\text{2D}}_{11} g^{\text{2D}}_{22}}$, we estimate a critical interspecies scattering length of $a_{12}^{c} = 66~a_0$. Since our experimental value $a_{12}>a_{12}^{c}$, the 2D mixture is likely to reside in the immiscible regime as in the 3D case. As shown in Fig. \ref{fig:2d} (c) and Fig. \ref{fig:2drad} (b), the total atom number of Na (Rb) is measured to be $1.7 \times 10^5$ ($5.2 \times 10^4$). The fitting of the low-density thermal wing with $\tilde{\mu}<0$ yields a chemical potential  $k_B \times$ 52.1 nK and a temperature 98.8 nK for Na atoms, and a chemical potential $k_B \times$ 52.0 nK for Rb atoms, where the two species are assumed to be in thermal equilibrium. We note that the measured temperature is significantly lower than the vertical vibrational energy scales, $\hbar\omega_z/k_B \approx 210~ (190)$ nK for Na (Rb), confirming that the gases are confined within the quasi-2D regime. The density profiles of thermal wings agree well with the self-consistent HFMF theory solutions. In the superfluid regions with $\tilde{\mu} > \tilde{\mu}_c$, we add another term into the local chemical potential $\mu_i(r) = \mu_i -V_i(r)- g_{ij}^{2\text{D}}n_j(r)$ to take into account the interspecies interactions, and then carry out similar self-consistent calculations to obtain the density in the superfluid region, which reasonably agrees with the experiments. Here we take $\tilde{\mu}_c \approx 0.22$ and $\tilde{g} \approx 0.156$ for Rb, and the dimensionless interspecies interaction $\tilde{g}_{12} =2m_{12}g^{\rm 2D}_{12}/\hbar^2\approx 0.08$. Notably, the Rb atoms form a dense central core due to their higher mass and compressibility in the steeper radial potential, while the Na atoms are partially expelled from the center, forming a partial ring. This signals the immiscible repulsion between the two atomic species in the quasi-2D geometry. Neglecting the effects of interspecies interactions \cite{Karle_2dmix_2019, Kobayashi_2dmix_2019}, we estimate the strictly 2D BKT transition temperatures for Na and Rb in the mixture to be approximately 257 nK and 111 nK, respectively. Since the measured temperature is comparable to these transition temperatures, finite-temperature effects likely contribute to the observed smearing of phase separation \cite{roy_2d-bose-mixture_2021}. A detailed analysis that accounts for the intermediate fluctuation regime is required in further theoretical studies.

\section{Conclusion and Outlook}

In conclusion, we have developed a new experimental apparatus capable of efficiently producing dual-species quantum degenerate mixtures confined in a quasi-2D geometry. We have detailed the design considerations for the vacuum system, along with the laser cooling and trapping procedures that collectively reduce atomic temperatures from room temperature down to the hundred-nanokelvin regime. The science chamber is engineered to accommodate various optical dipole traps and lattice geometries, magnetic fields for tuning interspecies Feshbach resonance between atoms near 347.6 G \cite{guo_scatter_2022}, and the precise installation of an in-vacuum electrode assembly to control ultracold NaRb molecules. Finally, high-resolution optical imaging enables observation of the \textit{in situ} density distributions of the quasi-2D Bose mixture. 

The 2D quantum mixture provides an ideal starting point for exploring a variety of intriguing ultracold low-dimensional phenomena. For instance, in a repulsively interacting and balanced mixture, one could tune the interspecies interaction strength or further reduce the mixture temperature to map the miscible-immiscible phase boundary \cite{Karle_2d-bose-mixture_2019, roy_2d-bose-mixture_2021} or investigate mixed bubbles \cite{Naidon_bose-mixture_2021}, potentially providing a route toward probing Andreev-Bashkin effects. Conversely, attractively interacting and balanced mixtures allow for the study of unique Bose droplet properties in low dimensions~\cite{hammer_droplets_2004, Petrov_droplet_2016, Pan_droplet_2022} and their finite-temperature behavior~\cite{Spada_droplet_2024}. For imbalanced 2D mixtures, our high-resolution imaging setup is well-suited to study Bose polarons in reduced dimensions and probe the induced quantum correlations between impurities and the baths~\cite{Ardila_polaron_2020, Colussi_lattice_polaron_2023, Comaron_impurity_2025}. Furthermore, the ultracold atomic mixture can be associated into $^{23}$Na$^{87}$Rb polar molecules~\cite{Guo_NaRb_2016}. With the in-vacuum electrode, these trapped 2D bosonic polar molecules can be polarized to study collision properties under shielding potentials engineered via uniform DC electric field~\cite{buechler_molphase_2007, Gorshkov_shield_2008}. 

\begin{acknowledgments}
We acknowledge the important discussions with Han-Ning Dai and Yi-Fei He. We thank Wei Gou and Bo-Yuan Wang for their early contributions to the construction of the apparatus and Zhen-Sheng Yuan for generous personnel support. This work was supported by the National Natural Science Foundation of China (under Grant No. 12274393, No.12241409, No. 12325407, No. 12488301), the Chinese Academy of Sciences, the Shanghai Municipal Science and Technology Major Project (Grant No. 2019SHZDZX01), and the Quantum Science and Technology-National Science and Technology Major Project (Grant No. 2021ZD0302101, 2021ZD0302001). 
\end{acknowledgments}

\textit{Data Availability} — The data that support the findings of this article are openly available \cite{data_link}.

\appendix

\setcounter{equation}{0}
\setcounter{figure}{0}

\renewcommand{\theequation}{A\arabic{equation}}
\renewcommand{\thefigure}{A\arabic{figure}}

\setcounter{section}{0}

\section{Modular 2D$^{+}$-MOT of $^{87}$Rb Atoms}
\label{app-Rb-2DMOT}

The Rb 2D-MOT chamber is an all-metal rectangular structure measuring 68 mm × 68 mm × 190 mm. On its four long sides,  custom-made rectangular fused silica windows are installed and sealed using indium wire cold pressing. A 45$^{\circ}$ inclined mirror, polished aluminum with a central through-hole, is mounted on the chamber's exit side to provide counter-propagating axial cooling light. The chamber is connected to the 3D-MOT chamber via a custom differential tube with an inner diameter of 3 mm and a length of 80 mm, yielding a conductance of 0.04 L/s. Owing to the low melting point of indium wires, the chamber was baked to a maximum temperature of around 110 $^{\circ}\mathrm{C}$. To supply the Rb vapor background, a glass ampoule containing 1 g of pure Rb metal was placed inside a brass tube during vacuum assembly and broken after baking. To maintain a sufficiently high partial pressure of Rb, a baffle flange was installed in front of a 20 L/s ion pump, effectively halving its pumping speed. After back-out, the chamber pressure reaches 1.1×10$^{-10}$ mbar as measured by the ion pump. 

\begin{figure}[h]
	\includegraphics[width=75mm]{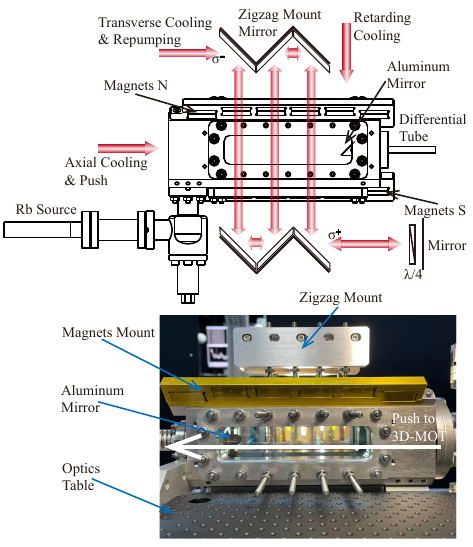}
	\caption{\label{fig:rb2d} Schematic and photograph of the $^{87}$Rb 2D$^{+}$ MOT. A polished aluminum mirror, beveled at a 45$^{\circ}$ angle, is mounted inside the chamber at the top of the differential pumping tube. This mirror reflects the axial cooling light to increase the Rb-atom flux. The lower image shows the assembled 2D-MOT chamber with one of the modular transversal-cooling optics mounted onto the threaded studs of an indium-sealed, elongated window.}
\end{figure} 

To generate the quadrupole magnetic field for the 2D-MOT, two sets of six NdFeB magnets are installed, as illustrated in Fig. \ref{fig:rb2d}. Each magnet measures 3 mm × 10 mm × 25 mm and has a magnetization of $8.7\times10^5$ A/m. On each side, the magnets are arranged into two groups: four inner magnets positioned 50 mm from the chamber center, and two outer magnets at 46.5 mm, with a uniform 8 mm spacing between adjacent magnets. Within a set (i.e., on the same side), all magnets share the same polarity, oriented perpendicular to their main surface. The two sets are mounted diagonally opposite with opposing relative polarities, producing a radial gradient of 11.2 G/cm for transverse laser cooling while maintaining a near-zero central field for axial molasses cooling and pushing.

To laser cool the atoms, the 2D-MOT cooling and repumping lasers are combined and then split into transverse and axial beams, as shown in Fig. \ref{fig:rb2d}. The Transverse beams are further divided into horizontal and vertical directions (1/e$^2$ beam diameter 22 mm) and reflected by custom mirrors mounted on a zigzag bracket. All four mirror brackets are attached directly to the threaded studs that seal the rectangular cooling windows, resulting in a compact and stable optical setup. Each mirror inside these brackets introduces a $\pi$ phase shift between the H and V polarizations with a precision better than $\pi$/100. These beams propagate back and forth within the chamber and are finally retro-reflected along the original paths, forming an elongated irradiation zone with appropriate $\sigma^{+}$/$\sigma^{-}$ polarizations required for laser cooling. The Axial beam is split into two counter-propagating beams (9.5 mm diameter) for molasses cooling, with mutually perpendicular linear polarizations. To enhance the atomic flux, a blue-detuned push beam (1.2 mm diameter) is applied. With optimized intensity and frequency parameters, a maximum loading rate of $^{87}$Rb 3D-MOT reaches 2×10$^{9}$ atoms per second--sufficient for most quantum gas experiments involving Rb atoms.

\section{Compact source of $^{23}$Na Atoms}
\label{app-Na-2DMOT}

A compact Na 2D-MOT uses an octagonal chamber with four CF35 vacuum windows on opposite sides spaced 135 mm apart \cite{lamporesi_na2dmot_2013}. An oven containing 10 g of pure sodium metal is mounted below the chamber. The central port of the chamber is connected to a CF35 four-way cross, which in turn connects to a getter pump, an all-metal angle valve, and a viewport for the push beam. The opposite side is connected to the 3D-MOT chamber via a 2.5 mm inner-diameter, 60 mm-long differential tube, resulting in a conductance of 0.05 L/s. 

During operation, the atomic oven is typically heated up to 215 $^{\circ}\mathrm{C}$. From the ideal gas law, the number density of Na atoms inside the oven is estimated to be approximately $5.5\times10^{10}\ \text{cm}^{-3}$, with an average atomic velocity of 670 m/s. The oven is connected to the chamber via an elongated CF16 reflux tube. The total effusive atomic flux through the tube aperture is given by
\begin{equation}
	\Phi = \frac{An}{4}\sqrt{\frac{8k_\mathrm{B} T}{\pi m}},
\end{equation}
which amounts to about 1.8×10$^{15}$ atoms$\cdot$s$^{-1}$. Here, A$\approx 200$ mm$^2$ is the effective area of the aperture, $n$ is the atomic density in the oven, $k_\mathrm{B}$ is the Boltzmann constant, $T$ is the thermodynamic temperature, and $m$ is the atomic mass. With the Na ampoule installed, the oven's expected service life exceeds 10 years.

\begin{figure}[h]
	\includegraphics[width=75mm]{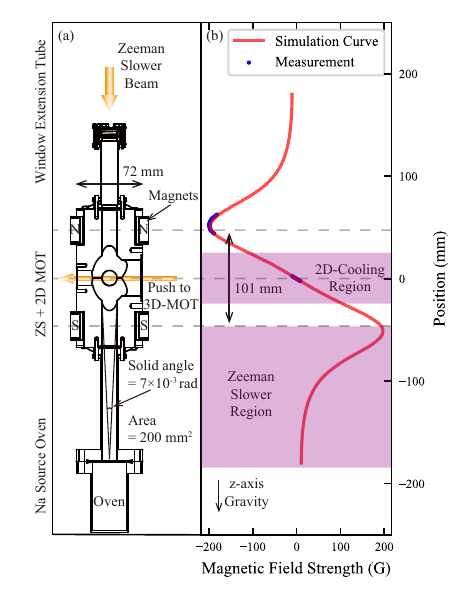}
	\caption{\label{fig:na2d} Schematic of the compact Zeeman slower for $^{23}$Na atoms. (a) Cross-sectional view of the Na 2D chamber, showing the permanent magnets that generate the required magnetic field  gradient for Zeeman slowing and transverse 2D-MOT cooling. (b) Simulated (red dots) and measured (blue dots) magnetic field strength for both Zeeman slower and 2D-MOT cooling.}
\end{figure}

To laser cool $^{23}$Na atoms, a quadrupole magnetic field is generated by four sets of NdFeB permanent magnets. Each set consists of a stack of nine magnets sharing the same polarity; the individual magnets have the same specifications as those used in the  Rb 2D-MOT. As illustrated in Fig. \ref{fig:na2d}(a), the left and right sets are spaced 72 mm apart and share the same polarity, whereas the upper and lower sets are spaced 101 mm apart and have opposite polarities. The magnetic field strength rises from nearly zero at the oven aperture to a maximum of 200 G, providing the deceleration field for a compact Zeeman slower. At the peak field, the Zeeman-slower cooling light is detuned by -360 MHz, corresponding to a capture velocity of 180 m/s. We note that this compact design is not efficient for heavier atoms such as Rb or Cs because their lower radiative deceleration rates would require a much longer slowing stage. Atoms pre-cooled by the Zeeman slower then enter a central magnetic-field gradient of 38 G/cm (along the vertical direction),  where they are subject to transverse 2D-MOT cooling.

In our setup, the repumping light for both Zeeman slowing and 2D-MOT cooling of $^{23}$Na is generated directly from the cooling laser using a resonant electro-optic modulator (EOM). The Zeeman-slowing light (14 mm diameter) is delivered through the upper CF16 viewport with linear polarization perpendicular to the quadrupole magnetic-field plane. Two separate 2D-MOT cooling beams (each with a 25 mm diameter) are sent through the tilted CF35 viewports and retro-reflected. A blue-detuned push light (1.3 mm diameter) directs the slowed atomic flux into the 3D-MOT chamber through the axial CF35 viewport. With this configuration, we achieved a maximal $^{23}$Na 3D-MOT loading rate of $\sim2.2\times10^8$ atoms s$^{-1}$\footnote{The Na atom number in the Dark-SPOT has been recently improved to $\sim1\times10^{9}$ atom s$^{-1}$ by carefully optimizing the Zeeman cooling beam size.}. 

\section{Dark-SPOT and Sub-Doppler Cooling}
\label{app-3DMOT}

Atoms of both species, after being pushed into the 3D-MOT chamber, are captured in a Dark-SPOT \cite{ketterle_dsmot_1993}. In many dual-species MOT setups, interspecies light-assisted collisions can lead to significant atom loss and heating, limiting the achievable density and overlap. The Dark-SPOT suppresses these losses and achieves a higher central density \cite{Ridinger_motloss_2011}. The quadrupole magnetic field is generated by a pair of coils mounted inside the CF100 re-entrant flanges, with an optimal axial ($z$-direction) gradient of 15 G/cm. Three pairs of counter-propagating, dual-color MOT beams were sent along three orthogonal directions. All cooling beams are locked to the red-detuned side of the $|F=2\rangle \rightarrow|F'=3\rangle$ cycling transition of the D2 line, where $F$ denotes the hyperfine energy level and $m_F$ its magnetic sublevel. The beam diameters are 37 mm for the 589 nm cooling beam and 43 mm for the 780 nm cooling beam. Repumping light is injected through one of the horizontal viewports, with a beam diameter of about 28 mm for both species. To increase the atomic density, a black aluminum spot is placed to block the central area of the repumping beam, creating an 8-mm-diameter dark zone at the center of the chamber. This allows atoms to accumulate in the $|F=1\rangle$ state. Other parameters such as laser intensities and detunings are summarized in Table \ref{tab:table2}. The peak optical depth (OD) of Na atoms in the dark-SPOT reaches 80, as determined by fitting the transmission curve of a frequency-scanned repumping probe beam. After a 3-second MOT loading phase, we can load $6.4\times10^8$ $^{23}$Na atoms with a temperature of 190 $\mu$K and 2.8×10$^9$ $^{87}$Rb atoms with 97 $\mu$K simultaneously.

\begin{table}[t]
	\caption{\label{tab:table2}%
		Optimized laser parameters for different cooling stages. Frequencies are reported relative to the target transition (e.g., $f_{2\rightarrow3}$ represents the frequency of the $|F=2\rangle \rightarrow|F'=3\rangle$ transition). Peak intensities are normalized to the saturation intensity of the $|F=2\rangle \rightarrow|F'=3\rangle$ transition.}
	\begin{ruledtabular}
		\begin{tabular}{ccc}
			&
			\textrm{Frequency} &
			\textrm{Intensity} \\
			\colrule
			\multicolumn{3}{c}{Na 2D-MOT}\\
			\colrule
			Na Zeeman Slower & $f_{2\rightarrow3}-36.8\ \Gamma_{\text{Na}}$\footnote{Natural Line Width of the D2-line $\ \Gamma_{\text{Na}} = 2\pi\cdot9.79$ MHz for $^{23}$Na} & 25.2 $I^{\text{Sat}}_{\text{Na}}$\footnote{Saturation Intensity $I^{\text{Sat}}_{\text{Na}} = 6.26$ mW/cm$^2$ for $^{23}$Na} \\
			Na 2D Cooling & 
			$f_{2\rightarrow3}-1.5\ \Gamma_{\text{Na}}$ & 2.6 $I^{\text{Sat}}_{\text{Na}}$ \\
			Na 2D Pushing&
			$f_{2\rightarrow3}+1.0\ \Gamma_{\text{Na}}$ & 1.3 $I^{\text{Sat}}_{\text{Na}}$ \\
			\colrule
			\multicolumn{3}{c}{Rb 2D-MOT}\\
			\colrule
			Rb Trans Cooling &
			$f_{2\rightarrow3}-2.6\ \Gamma_{\text{Rb}}$\footnote{Natural Line Width of the D2-line $\ \Gamma_{\text{Rb}} = 2\pi\cdot6.07$ MHz for $^{87}$Rb} &  15 $I^{\text{Sat}}_{\text{Rb}}$\footnote{Saturation Intensity $I^{\text{Sat}}_{\text{Rb}} = 1.67$ mW/cm$^2$ for $^{87}$Rb} \\
			Rb Trans Repumping &
			$f_{1\rightarrow2}$ &  4 $I^{\text{Sat}}_{\text{Rb}}$ \\
			Rb Axial Cooling &
			$f_{2\rightarrow3}-2.6\ \Gamma_{\text{Rb}}$ &  2.2 $I^{\text{Sat}}_{\text{Rb}}$ \\
			Rb Retarding Cooling & $f_{2\rightarrow3}-2.6\ \Gamma_{\text{Rb}}$ &  1.2 $I^{\text{Sat}}_{\text{Rb}}$ \\
			Rb 2D Pushing & $f_{2\rightarrow3}+1.2\ \Gamma_{\text{Rb}}$ & 10 $I^{\text{Sat}}_{\text{Rb}}$ \\
			\colrule
			\multicolumn{3}{c}{3D-MOT}\\
			\colrule
			Na 3D-MOT Cooling & $f_{2\rightarrow3}-2.0\ \Gamma_{\text{Na}}$ & 3.5 $I^{\text{Sat}}_{\text{Na}}$ \\
			Na Dark Spot Repumping & $f_{1\rightarrow2}$ & 0.7 $I^{\text{Sat}}_{\text{Na}}$ \\
			Rb 3D-MOT Cooling & $f_{2\rightarrow3}-3.3\ \Gamma_{\text{Rb}}$ & 16 $I^{\text{Sat}}_{\text{Rb}}$ \\
			Rb Repumping & $f_{1\rightarrow2}$ & 3.4 $I^{\text{Sat}}_{\text{Rb}}$ \\
		\end{tabular}
	\end{ruledtabular}
\end{table} 

We empirically found that directly loading the 3D-MOT atoms into the magnetic trap was inefficient. To address this issue, a sub-Doppler cooling stage is essential for increasing the atomic PSD. For Rb atoms, we applied standard red-detuned molasses cooling by increasing the cooling detuning to 80 MHz and reducing its power to 90\%  in a fully compensated ambient magnetic field. After 5 ms of sub-Doppler cooling, the temperature of the Rb atomic cloud dropped to 54 $\mu$K. More importantly, the number of Rb atoms captured in the magnetic trap increased by a factor of 29 compared to the case without molasses cooling. For Na atoms, we employed a D2-line gray molasses cooling scheme \cite{shi_nagray_2018}, which relies on a $\Lambda$-type three-level system using the hyperfine levels of $|F=1\rangle$ and $|F=2\rangle$. The cooling light for the gray molasses is the same as the 3D-MOT cooling light, while the corresponding repumping light is generated via a resonant EOM. We found the optimal single-photon detuning for both beams to be +2.8 $\Gamma$ relative to the $|F'=2\rangle$ excited state, with an optimal intensity ratio of 5:1. After a 5 ms gray molasses stage, the Na atom temperature reached 36 $\mu$K, and the number of atoms loaded into the magnetic trap doubled relative to the case without this process. After molasses cooling, we extend the cooling light by 0.1-1 ms longer than the repumping light to depump all the atoms into the F=1 manifolds.

\section{Magnetic Trap Loading and Evaporation}
\label{app-evaporation}

Forced evaporative cooling was then carried out in the magnetic trap to further increase the PSD. Following the molasses stage, the magnetic field was rapidly ramped up to above 30 G/cm to levitate both species against gravity, where only atoms in the $|F=1,m_F=-1\rangle$ sublevel (1/3 of total atoms) were loaded into the magnetic trap. Then we slowly ramped to 160 G/cm over 60 ms. This further compresses the atomic ensemble, resulting in temperatures of $\sim200\mu$K and PSDs on the order of $10^ {-6}$ for both species. For RF-forced evaporation, we installed a square macro-ceramic (26×40 mm) inside the 3D-MOT chamber and wound six turns of Kapton-insulated copper wire. The assembly was positioned 47 mm from the magnetic trap center. An external RF signal, at an input power of 33 dBm, drove the atomic spin-flip transitions. Both Na and Rb atoms are held in the magnetic trap in the $|F=1, m_F=-1\rangle$ state, and their Zeeman splittings are nearly identical under low field. Consequently, a single frequency-swept RF signal simultaneously evaporates atoms of both species, which is another feature of this work compared to earlier experiments with Na-Rb mixture \cite{wang_bec_2016, rosenberg_molecule_2022}.

\begin{figure}[t]
	\includegraphics[width=75mm]{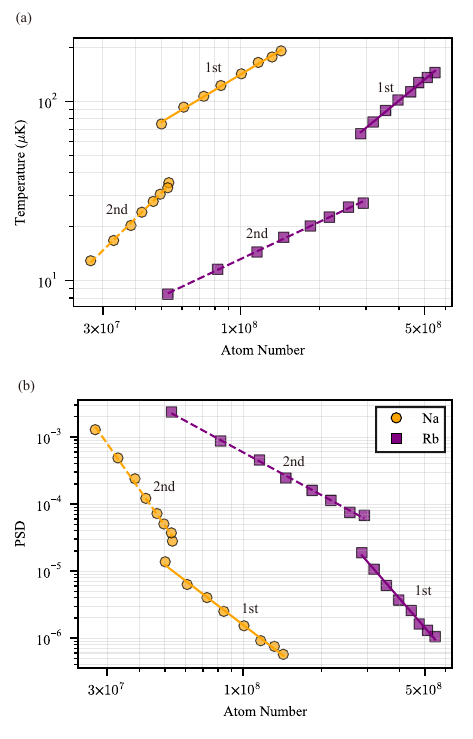}
	\caption{\label{fig:rfevap} Simultaneous RF evaporation of $^{23}$Na and $^{87}$Rb atoms in a quadrupole magnetic trap. (a) Temperature as a function of atom number. (b) PSD as a function of atom number. Solid lines correspond to the first RF evaporation stage performed at a magnetic gradient of 160 G/cm;  dashed lines correspond to the second stage at a reduced gradient of 64 G/cm.}
\end{figure}

At a magnetic trap gradient of 160 G/cm, the RF frequency was linearly swept from 30 MHz down to 10 MHz over 3 s, after which the RF field was turned off. The magnetic gradient was then adiabatically lowered to 64 G/cm within 500 ms to suppress three-body inelastic losses at high densities. Subsequently, the RF field was switched back on, and its frequency was further swept from 7 MHz to 2 MHz over 3 s. As RF evaporation in the magnetic trap proceeds, the atomic temperature falls, and the PSD rises. For a quadrupole magnetic trap, the PSD reads
\begin{equation}
	\Lambda_{\text{QT}} = \frac{N( \beta \mu  B' \lambda_{\text{dB}})^3 }{32\pi},
\end{equation}
where $ N $ is the total atom number, $ \mu $ is the atomic magnetic moment, $ B' $ is the magnetic trap gradient along $z$ direction, $ \beta = 1/(k_\text{B} T) $, and $\lambda_{\text{dB}} = \hbar \sqrt{\frac{2\pi  }{m k_\text{B} T}}$ is the atomic de Broglie wavelength. Fig. \ref{fig:rfevap} shows the results of a typical dual-species RF evaporation in the quadrupole magnetic trap. After two stages of RF evaporation, we obtained 2.7×10$^7$ $^{23}$Na atoms at a temperature of 12.9 $\mu$K with a PSD of 1.3×10$^{-3}$, and 5.3×10$^7$ $^{87}$Rb atoms at a temperature of 8.4 $\mu$K with a PSD of 2.4×10$^{-3}$. The evaporation efficiency $\beta = \partial \ln(T)/\partial \ln(N)$ and  $\Gamma = -\partial \ln(\Lambda)/\partial \ln(N)$ were extracted for both species. Fitting gives $\beta_\text{Na}=0.88$, $\Gamma_\text{Na}=3.28$ and $\beta_\text{Rb}=1.21$, $\Gamma_\text{Rb}=5.45$ for the first evaporation stage, and $\beta_\text{Na}=1.44$, $\Gamma_\text{Na}=5.49$ and $\beta_\text{Rb}=0.69$, $\Gamma_\text{Rb}=2.13$ for the second stage. 

\begin{figure}[t]
	\includegraphics[width=80mm]{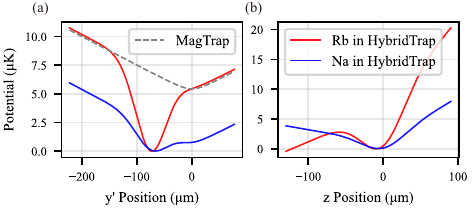}
	\caption{\label{fig:hybrid} Potential energy profiles of Na and Rb atoms in the hybrid trap near the end of evaporation toward quantum degeneracy. (a) Horizontal potential along the transverse direction of the optical trap. The minimum lies near 75 $\mu$m, and the trap depth for Rb is considerably larger than for Na along this axis. (b) Total vertical potential for both species. In a pure 1064 nm optical trap, the depth for Na atoms is only about $\sim$0.31 times that for Rb. In the hybrid trap configuration, however, magnetic levitation effectively increases the trap depth for Na while lowering that for Rb atoms, thereby enabling efficient evaporation of Rb atoms.}
\end{figure}

As the temperature continued to decrease, a 1064 nm laser beam was introduced to suppress the Majorana spin-flip losses near the magnetic zero point. This single beam, later detailed as the transport optical trap, has a beam waist of 55 $\mu$m. Its waist was aligned vertically with the magnetic trap center, with a horizontal offset of about 75 $\mu$m \cite{wang_bec_2016}. During the final 0.4 s of the second RF evaporation stage,  the beam was turned on to a trap depth of $k_\text{B} \times$13 $\mu$K for Rb ($4~\mu$K for Na), thereby forming a hybrid magnetic-optical trap. Typically, the evaporation of Rb atoms in this optical trap is rather inefficient due to its large polarizability at the trap light's wavelength. As a result, Rb atoms are mainly sympathetically cooled by evaporating Na atoms in the optical trap (as in the crossed-ODT of the science chamber). Here, we conceived a new strategy to overcome this constraint and efficiently evaporated both species, thereby realizing dual-species quantum degeneracy in this hybrid trap. To do so, after completing two stages of RF evaporation in the quadrupole trap, we adiabatically raised the optical-trap depth to 114 $\mu$K for Rb (43~$\mu$K for Na) within 100 ms while simultaneously reducing the vertical magnetic gradient to 15 G/cm. The remaining horizontal magnetic gradient provides the necessary axial confinement for the mixture in the hybrid trap. Moreover, the vertical gradient was set to levitate Na atoms against gravity but below the levitation threshold for Rb. This configuration allows independent control over the effective trap depths for Na and Rb in a shallow ODT. The total potential of the hybrid trap is given by \cite{lin_bec_2009},
\begin{equation}
\begin{aligned}
	U(x', y', z) &= \mu B' \sqrt{\frac{x'^2}{4} + \frac{y'^2}{4} + z^2}\\ &- U_{\text{ODT}} \exp{[- 2 \frac{(y'-y'_0)^2 + z^2}{\omega_0^2} ]} - m\text{g} z,
\end{aligned}
\end{equation}
with $x'$ and $y'$ the coordinates in the 3D-MOT chamber (see Fig. \ref{fig:sciopt}), $U_{\text{ODT}} $ the depth of the ODT potential, $\omega_0 \approx 55\ \mu\text{m}$ the trap beam waist, $y'_0=-75~\mu$m the horizontal position offset of the beam center relative to the center of the magnetic trap, and $\text{g}$ the gravitational acceleration.

After holding the atoms for 400 ms, the optical trap depth was exponentially ramped down from 114 $\mu$K to 7.5 $\mu$K (for Rb) during 4 s with a time constant of 500 ms. During this stage, the effective depth for Rb in the vertical direction is gradually reduced by gravity (see Fig. \ref{fig:hybrid}), which facilitates the sympathetic cooling of Na atoms. The levitation gradient can also be adjusted to control the trap depth of Na. Without the magnetic levitation, however, the trap depth for Rb would be much larger than that for Na, resulting in sympathetic cooling in which Rb is cooled primarily by evaporating Na. At the end of hybrid trap evaporation, the measured trap frequencies were $\omega_{y'}$ = 2$\pi$× 152 (149) Hz radially and $\omega_{x'}$ = 2$\pi$× 58 (29) Hz axially for Na (Rb). Fitting time-of-flight (TOF) images with a Thomas-Fermi profile gave atom numbers of 2.8×10$^5$ (6.6×10$^5$) and condensation fractions of 51\% (47\%) for the two species, respectively. The evaporation efficiencies were $\beta= 2.42$ for Na and $\beta=1.53$ for Rb. Our independently optimized 2D-MOTs enable efficient simultaneous RF evaporation of both species, while the hybrid trap allows independent tuning of Rb atom numbers to produce balanced dual-species BECs.

\bibliography{setup_bibref}

\end{document}